\documentclass[aip,superscriptaddress,reprint]{revtex4-1}%
\usepackage{epsfig,pslatex,latexsym,times,amssymb,amsmath,graphicx}
\usepackage{bm, float, lipsum}
\usepackage{amsmath}
\usepackage{amsfonts}
\usepackage{amssymb}
\usepackage{mathrsfs}
\usepackage{color}
\usepackage{graphicx}%
\usepackage{tikz}

\providecommand{\U}[1]{\protect\rule{.1in}{.1in}}
\begin{document}
\author{N. J. Harmon}
\affiliation{Department of Physics and Astronomy and Optical Science and Technology Center, University of Iowa, Iowa City, Iowa
52242, USA}
\author{M. Wohlgenannt}
\affiliation{Department of Physics and Astronomy and Optical Science and Technology Center, University of Iowa, Iowa City, Iowa
52242, USA}
\author{M. E. Flatt\'e}
\affiliation{Department of Physics and Astronomy and Optical Science and Technology Center, University of Iowa, Iowa City, Iowa
52242, USA}
\affiliation{Department of Applied Physics, Eindhoven University of Technology, P.O. Box 513, 5600 MB, Eindhoven, The Netherlands}
\date{\today}
\title{Manipulation of the electroluminescence of organic light-emitting diodes via fringe fields from patterned magnetic domains}  
\begin{abstract}
We predict very large changes in the room-temperature electroluminescence of thermally-activated delayed fluorescence organic light emitting diodes near patterned ferromagnetic films. These effects exceed the changes in a uniform magnetic field by as much as a factor of two. We describe optimal ferromagnetic film patterns for enhancing the electroluminescence. A full theory of the spin-mixing processes in exciplex recombination, and how they are affected by hyperfine fields, spin-orbit effects, and ferromagnetic fringe field effects is introduced, and is used to describe the effect of magnetic domain structures on the luminescence in various regimes. This provides a method of enhancing light emission rates from exciplexes and also a means of efficiently coupling information encoded in magnetic domains to organic light emitting diode emission.
\end{abstract}
\maketitle

Large changes in conduction and luminescence from modest ($<100$~Oe) magnetic fields\cite{Kalinowski2003, Mermer2005b, Prigodin2006, Desai2007} or from patterned ferromagnetic films\cite{Wang2012, Macia2014} have been observed in organic light-emitting diodes\cite{Friend1999} (OLEDs) at room temperature.   
Emission intensity changes in excess of an order of magnitude have been found recently when a uniform magnetic field is applied to emitting regions consisting of co-evaporated blends that emit via exciplex recombination.\cite{Wang2016}
The mechanism of these effects is often the manipulation of  spin mixing or spin-dependent recombination of precursor pairs. If the singlet-triplet energy splitting is small, such as in highly-efficient thermally-activated delayed fluorescence (TADF) devices (with exciplex emission), then direct spin mixing of the exciplexes is possible. Experimental measurements and theoretical calculations have identified the dominant spin mixing mechanism in the most common TADF materials to be due to the difference in Land\'e $g$~factor between the electron and hole that recombine. Spin mixing driven by this mechanism only effectively mixes one of the three triplet eigentstates with the singlet, leading to an expected maximum efficiency improvement of a factor of $2$.\cite{Wang2016} The remaining two triplet eigentstates are not efficiently harvested into the singlet channel by the applied magnetic field.

We predict the ability to harvest all the triplet eigenstates using the fringe field from a nearby ferromagnetic film. Thus the enhancement of OLED emission would be a factor of $4$ over the zero-field value, and a factor of $2$ over the emission when a uniform magnetic field is applied. The ability to mix the other two triplet eigenstates into the singlet channel occurs because of the magnetic field gradient that is generated by patterned ferromagnetic films. When other spin-mixing mechanisms are present, fringe fields from remanent magnetic states act as a means to either boost or reduce light emission from those mechanisms. By calculating the fringe field from various patterned ferromagnetic film configurations we predict that even OLED active regions that are 100 nm away from the ferromagnetic film can achieve the maximal (factor of 4) enhancement of light emission.

\begin{figure*}[ptbh]
 \begin{centering}
        \includegraphics[scale = 0.373,trim = 0 0 0 0, angle = -0,clip]{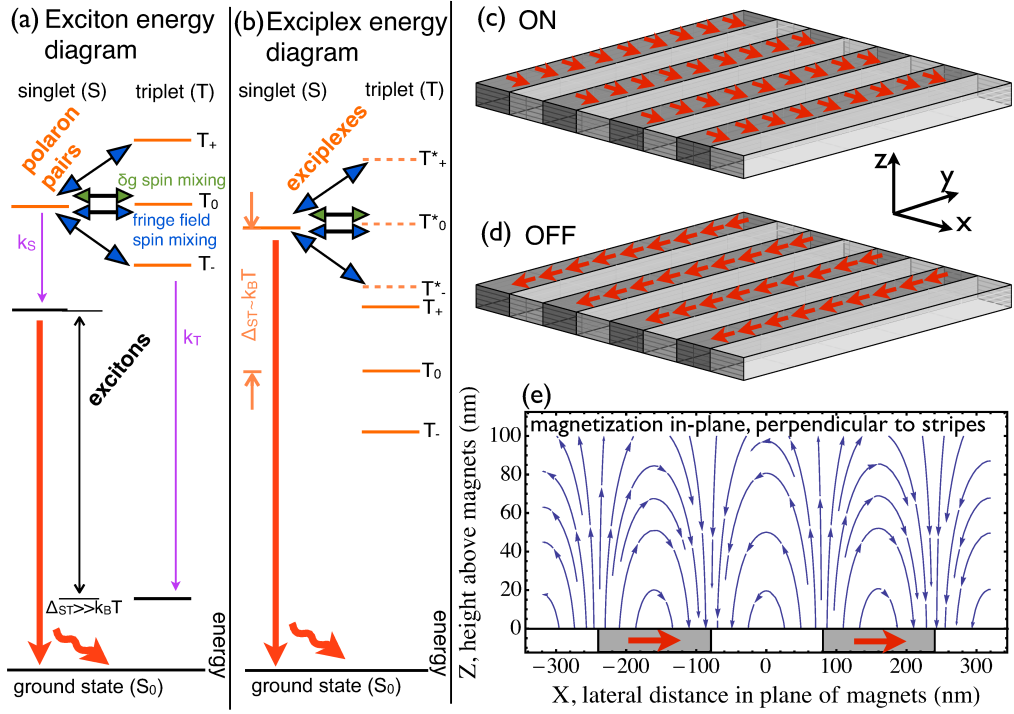}
        \caption[]
{
(a) Diagram of exciton energies and recombination pathway. 
(b) Diagram of exciplex energies and recombination pathway.
(c) and (d) depict, respectively, the ON and OFF domain configurations described in the text.
(e) In-plane magnetic configuration (c) with fringe fields directions (arrows only depict directions and not magnitudes) in the $x-z$ plane. Parameters used: $t = 20$ nm, $M_s = 8 \times 10^5$ A/m, and $a = 160$ nm. The calculation assumes $i_{max}\gg 1$.
}\label{fig:Fig1}
        \end{centering}
\end{figure*}

We now proceed to describe the effect of a spin mixing process on light emission in detail, as a prelude to describing the new form of spin mixing provided by the ferromagnetic fringe fields. In the typical OLEDs studied, electron and hole polarons encounter one another in the bulk and temporarily form loosely bound states, termed polaron pairs, with a 3:1 ratio of triplet to singlets.
Polaron pairs proceed further to combine into excitons [Figure \ref{fig:Fig1}(a)]; whether the precursors are singlet or triplet influences the rate of the exciton formation.
Only singlet excitons lead to significant electroluminescence, and thus 
most organic semiconductors are fluorescent materials with an internal electroluminescence quantum efficiency limited to 25\%.\cite{Rothberg1996} 
Magnetic fields influence the polaron pair formation kinetics, as interactions that cause singlet-triplet intersystem crossing, or spin mixing, lead to changes in the exciton singlet/triplet formation ratios. The large ($\sim 1 $~eV) singlet-triplet exchange energy for excitons precludes further spin mixing.
Magnetic field effects on co-evaporated donor/acceptor organic blends can follow different dynamics [Figure \ref{fig:Fig1}(a)], as light emission occurs not through an exciton (or intramolecular) pathway but through an exciplex (or intermolecular) route. These materials draw interest for OLEDs because they allow  triplet exciplexes to convert to emissive singlets through spin mixing, as the  ($< 100$ meV) singlet-triplet energy difference is provided by thermal excitation. The emission from the triplet-to-singlet up-conversion is known as thermally assisted delayed fluorescence (TADF) since the emission increases with temperature.\cite{Goushi2012,Uoyama2012}

The origins of spin mixing, which leads to the above magnetic-field effects on the conductance and luminescence, include the hyperfine interaction (HF) between polarons and hydrogen nuclei which are omnipresent in most organic systems\cite{Bobbert2007, Harmon2012a}, as well as from differences in Lande $g$-factors ($\delta g$) between the positive and negative polarons.\cite{Wang2008, Schellekens2011, DevirWolfman2014, Wang2016}
Investigations of magnetically-switched OLED light emission led to the key demonstrations that light emission  can also be controlled by nearby ferromagnetic films \cite{Wang2012, Macia2013, Macia2014}, which provide alternate and sometimes stronger mechanisms of spin mixing.
Electrically isolated thin magnetic films below an OLED modify the charge transport and light emission  dramatically  when magnetic fringe fields penetrate the OLED. 
We examine  patterned domain stripes, Fig.~1(c-e), which are easier to theoretically study then the random domains of Refs.~\onlinecite{Harmon2013b, Macia2014}.
By appropriately orienting these domains, spin mixing via fringe fields can be switched on and off.  Additionally, the form of the fringe field interaction allows for unprecedented large changes in the electroluminescence $\Delta  \text{EL}/ \text{EL} = ( \text{EL}(ON) -  \text{EL}(OFF))/ \text{EL}(OFF) = 300 \%$.
Lastly, by adjusting the spacer width that separates the magnet from the organic layer, we  predict how future measurements  could distinguish the microscopic origins of exciplex spin dynamics.

We begin with the theoretical description of the charge and spin dynamics of exciton and exciplex recombination shown in Figure \ref{fig:Fig1}(a,b). In the well-studied  exciton scheme (1) polarons form singlet and triplet polaron pairs at rates $G_S$ and $G_T$ (2) spin mixing occurs between the polaron pair states (3) meanwhile polaron pairs collapse into excitons at respective rates $k_S$ and $k_T$ (4) spin evolution of excitons ceases due to large exchange energy and finally (5) singlet excitons luminesce. There is also a possibility for polaron pairs to dissociate at a rate $k_D$ though this can often be neglected.\cite{Kersten2011a, Macia2014} 
Early studies of exciplexes have assumed a very similar picture to the of excitons where exciplexes simply play the role of exciton. \cite{Crooker2014, Liu2014, Hontz2015, Ling2015}
Spin does not evolve in the exciplex state since the exchange splitting, though smaller than in excitons, is still as large as and more often larger than room temperature.
The magnetoelectroluminescence values achieved $\sim10 \%$ are consistent with HF spin mixing and $k_S \lesssim k_T$.
On the other hand, other workers modeled the situation as in Figure \ref{fig:Fig1}(b) where the polaron pair formation is skipped or of less importance and spin mixing occurs primarily within exciplexes.\cite{Basel2015, Wang2016, Lei2016}
Near room temperature an activated behavior is observed in the magnetoelectroluminescence which suggests that spin mixing occurs between singlets and some excited triplet states (T*) that lie near in energy to the singlets.
The magnetoelectroluminescence measurements of Ref. \onlinecite{Wang2016}, which are the among the highest recorded ($> $60 \%), also rule out HF in favor of $\delta g$ spin mixing. 
The successes of the two models of exciplex magnetoelectroluminescence suggest to us that different coevaporated blends may exhibit different routes to light emission.
In addition to providing larger luminescence, we propose that by allowing for fringe field spin mixing, light can be shed on the microscopic processes that lead to radiative recombination for exciplexes.

Consider a thin magnetic film with striped domains that emanates fringe fields in the organic layer. Two options for striped patterns are shown in Figure \ref{fig:Fig1}(c,d); the repeats occur every $a$ along the $x$-axis while the film extends far out in the $y$-direction to $\pm c$. The magnetic films have a thickness, $t$. We expect that domains magnetized perpendicular to the plane to be hardest to manufacture due to the magnetic anisotropy. 
Figure \ref{fig:Fig1}(c) is possible with sufficiently large $a$ and applied field to set the domains perpendicular to the stripes. As shown below, fringe fields appear in this orientation so this configuration is defined to be ON. Figure \ref{fig:Fig1}(d) is the most energetically favorable configuration but no fringe fields are produced if edge effects are neglected. The absence of fringing fields leads us to define this magnetic state as OFF. 
We define a figure of merit to be $\Delta  \text{EL}/ \text{EL}$. It should be noted that applied fields are not necessarily present when determining $\Delta  \text{EL}/ \text{EL}$ since the domains can be oriented with a `set' applied field and remain in that state after the field is switched off.\cite{Macia2013, Macia2014}
In fact these remanent magnetizations are what we are interested in so we can safely ignore any applied field when using our figure of merit. Of course an external field is needed to switch the domains between ON and OFF.

The magnetic scalar potential from a ferromagnet with magnetization, $\bm{M}$, volume, $V$, and surface, $S$, as shown in Figure \ref{fig:Fig1}(c) is\cite{Jackson1998}
\begin{equation}
\Phi(\bm{R}) = \sum_{i} \frac{1}{4 \pi}M_{s, i} \int_{S_i} \frac{\bm{n}_i \cdot \bm{M}_i(\bm{r}_i)}{|\bm{R} - \bm{r}_i |}dA_i,
 \end{equation}
where each domain is uniformly magnetized and denoted by an index $i$.
$\bm{R} = (X, Y, Z)$ is the position outside the magnet, the indexed $\bm{r} = (x, y, z)$ denote position within the magnet, and $A_i$ are area elements of the magnet's surface.
Since $\bm{H} = \bm{B}/\mu_0$ outside a magnetized volume, $\bm{B} = -\mu_0 \nabla \Phi(\bm{R})$ where $\mu_0 = 4 \pi \times 10^{-7}$ N/A$^2$.
We make the following assumptions: $M_{s, i}$ and the length and width of $S_i$  are all constant for all $i$ in a given configuration of the magnet. Simple modifications to these assumptions such as alternating saturated magnetization, $M_s$, can be handled if needed.

As shown in Fig. \ref{fig:Fig1}(e), the magnetized domains are separated spatially by non-magnetic stripes. Magnetic surface charge densities form on each domain wall and alternate between + and -. These surfaces are located at each $x_i$ and have an area $2 c (z_t - z_b) = 2 c t$ where $z_t$ and $z_b$ are the top and bottom positions of the magnet. We define $x = 0$ to be halfway in between two such oppositely `charged' plates (negative/positive plate lies at $x = \pm a/2$). Each plate (or actually domain wall) is then indexed by $x_i = (i + \frac{1}{2})a$ with $-i_{max} < i < i_{max} - 1$. The $x$-edge length is $L_x = (2 i_{max} -1)a$.
The surface integrations can be found in closed form solution though they are prohibitively expansive.
This motivates us to find the simpler limiting functional form of our derived $\bm{B}_{\text{FF}}(\bm{R})$ as a function of $Z$. We make an approximation for infinite stripes (\emph{i.e.} $c\rightarrow \infty$) that $t \ll a, Z$ which allows us to write reduced expressions for the fringe fields after expanding each term in small $t$:
 \begin{eqnarray}\label{eq:approx}
&&\bm{B}_{\text{FF}}(\bm{R}) =\\
{} &&- \frac{\mu_0}{4 \pi}M_{s} \frac{4 \pi  t }{a}\frac{\{-\cos \left(\frac{\pi  X}{a}\right) \cosh \left(\frac{\pi
    Z}{a}\right),~0, ~\sin
   \left(\frac{\pi  X}{a}\right) \sinh \left(\frac{\pi  Z}{a}\right)\}}{\cos \left(\frac{2 \pi  X}{a}\right)+\cosh
   \left(\frac{2 \pi  Z}{a}\right)}.\nonumber
 \end{eqnarray}
Figure \ref{fig:Fig1}(e) shows the direction of the fringe fields above the magnetic domains in the $x-z$ plane.

The Hamiltonian of the polaron pair (for exciton model) or exciplex is 
$\mathscr{H} = \mathscr{H}_{HF} + \mathscr{H}_{FF}$
where
 \begin{equation}
 \mathscr{H}_{\text{HF}} = 
 \frac{g_1+g_2}{2} \mu_B  \Big(\bm{B}_{\text{HF}}(\bm{r}_1) \cdot \bm{S}_1 + \bm{B}_{\text{HF}}(\bm{r}_2) \cdot \bm{S}_2\Big),
\end{equation}
 \begin{eqnarray}
\mathscr{H}_{\text{FF}} &=& 
 \frac{g_1+g_2}{2} \mu_B\frac{\bm{B}_{\text{FF}}(\bm{r}_1) +\bm{B}_{\text{FF}}(\bm{r}_2) }{2} \cdot (\bm{S}_1  + \bm{S}_2) +  {}  \\
 {}&&   \frac{g_1+g_2}{2} \mu_B\frac{\bm{B}_{\text{FF}}(\bm{r}_1)  - \bm{B}_{\text{FF}}(\bm{r}_2) }{2} \cdot (\bm{S}_1  - \bm{S}_2)   ;\nonumber 
 \end{eqnarray}
where $\bm{B}_{\text{HF}}$ is the hyperfine field, $\bm{r}_{1,2}$ ($\bm{S}_{1,2}$ ) are the positions (spins) of the two polarons, and $\delta_g = g_1 - g_2$. 
Since for typical values of $\delta g$ ($< 10^{-3}$), the fringe field mechanism swamps the $\delta g$ mechanism, we set $\delta g = 0$.

The spin dynamics of either the polaron pair or exciplex are encapsulated within a two-spin density matrix $\rho$ which evolves according to the stochastic Liouville equation: \cite{Haberkorn1976, Schellekens2011, Kersten2011a, Macia2014, Wang2016}
\begin{equation}\label{eq:SLE}
\frac{\partial \rho}{\partial t} =  - \frac{i}{\hbar} [\mathscr{H},  \rho] - \frac{1}{2} \{  k_S P_S + k_T P_T, \rho \} - k_D \rho + \frac{1}{4}\hat{G}
\end{equation}
where $P_S$ and $P_T$ are the singlet and triplet projection operators, $\text{Tr}\rho \ll 1$, and $\hat{G}$ is the exciplex generation matrix. For exciton model, it is assumed that $G_S = G_T$ while $G_T$ in the exciplex model considers only those exciplexes that are activated near the singlet level so $G_T \leq G_S$. The exciplex effects are maximal for $G_T = G_S$ which for simplicity we assume throughout our calculations here. This assumption corresponds to $k_B T \gg \Delta_{ST}$.

To perform our calculation of the figure of merit, Eq. (\ref{eq:SLE}) is solved in steady state with $\bm{B}_{\text{FF}}(\bm{r}_1)$ being calculated in the organic layer from Eq. (\ref{eq:approx}) where $\bm{r}_1$ is determined randomly  within a box of height $30$ nm, positioned at $Z = Z_{min}$ above the magnet, and lateral size equal to the magnetic film's.
Since the current path is in $\hat{z}$, the average carrier hop occurs in the $z$-direction so $\bm{B}_{\text{FF}}(\bm{r}_2) = \bm{B}_{\text{FF}}(\{ x_1, y_1, z_1 + d  \})$ where $d$ is the hopping length. We fix $d = 1$ nm.
If transport occurred laterally, we expect fringe field influence to be smaller since fringe fields do not change exponentially in $X$ but sinusoidally. 
After solving for $\rho$, the EL is determined from $\text{EL} \propto k_S \text{Tr} P_S \rho$.

\begin{figure}[ptbh]
 \begin{centering}
        \includegraphics[scale = 0.35,trim = 0 0 0 0, angle = -0,clip]{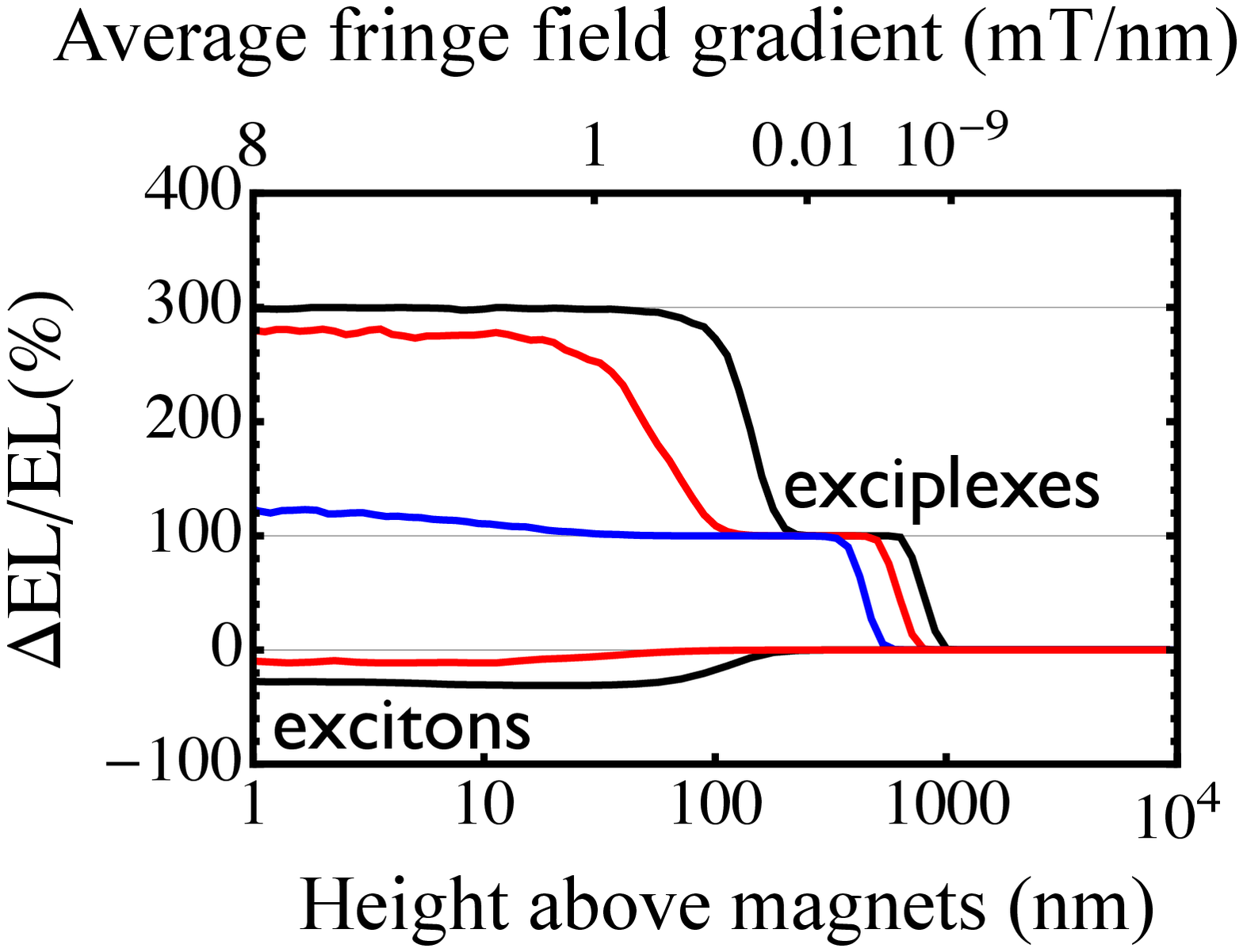}
        \caption[]
{Change in EL versus average fringe field gradient for exciplexes: $k_D = 10^{-12}$ ns$^{-1}$ (black), $k_D = 10^{-9}$ ns$^{-1}$ (red), $k_D = 10^{-6}$ ns$^{-1}$ (blue), and $k_S = 3 \times 10^{-3}$ ns$^{-1}$, $k_T = 0$ for all and excitons: black: $k_S = 3 \times 10^{-2}$ ns$^{-1}$, $k_T = 6 \times 10^{-2}$ ns$^{-1}$; red: $k_S = 3 \times 10^{-1}$ ns$^{-1}$, $k_T = 6 \times 10^{-1}$ ns$^{-1}$, and $k_D = 0$ for both. The patterned FM is permalloy with $M_s = 8 \times 10^{5}$ A/m, $t = 20$ nm, and $a = 160$ nm. The hyperfine field is taken to be zero.
}\label{fig:MEL}
        \end{centering}
\end{figure}

Now we discuss the chief results from our calculations. Figure \ref{fig:MEL} shows $\Delta \text{EL}/ \text{EL}$ under the operation of only the fringe field ($\mathscr{H}_{HF}=0$) for both the exciton and exciplex pictures.
Beginning on the right, if there is no fringe field gradient there is no spin mixing or change in EL.
As the height above the magnets, $Z_{min}$, is reduced (and gradient increased), the figure of merit plateaus at 100\%.
The reason for this plateau is reminiscent of the $\delta g$ mechanism for which all $T_0$ states may upconvert to $S$ states, doubling the EL.\cite{Wang2016}
It indicates that spin mixing between $T_{\pm}$ and $S$ states are much slower compared to $T_0$ and $S$ states.
$k_D$ is large enough to prohibit further mixing which explains why the curves shift left as $k_D$ is increased.
As the gradient increases further and dissociation is sufficiently small, the figure of merit rises to 300\% which is indicative of all triplet exciplexes upconverting and recombining as singlets.
Since the spin-selective rates for excitons vary little, contrary to what occurs with exciplexes, the figure of merit takes on less extreme values. We assume $k_S \lesssim  k_T$. \cite{Liu2014}
Alternatively, if  $k_S \gtrsim  k_T$ (not shown), the excitonic $\Delta  \text{EL}/  \text{EL}$ changes sign but still does not reach the magnitude seen for exciplexes.
\begin{figure}[ptbh]
 \begin{centering}
        \includegraphics[scale = 0.35,trim = 0 0 0 0, angle = -0,clip]{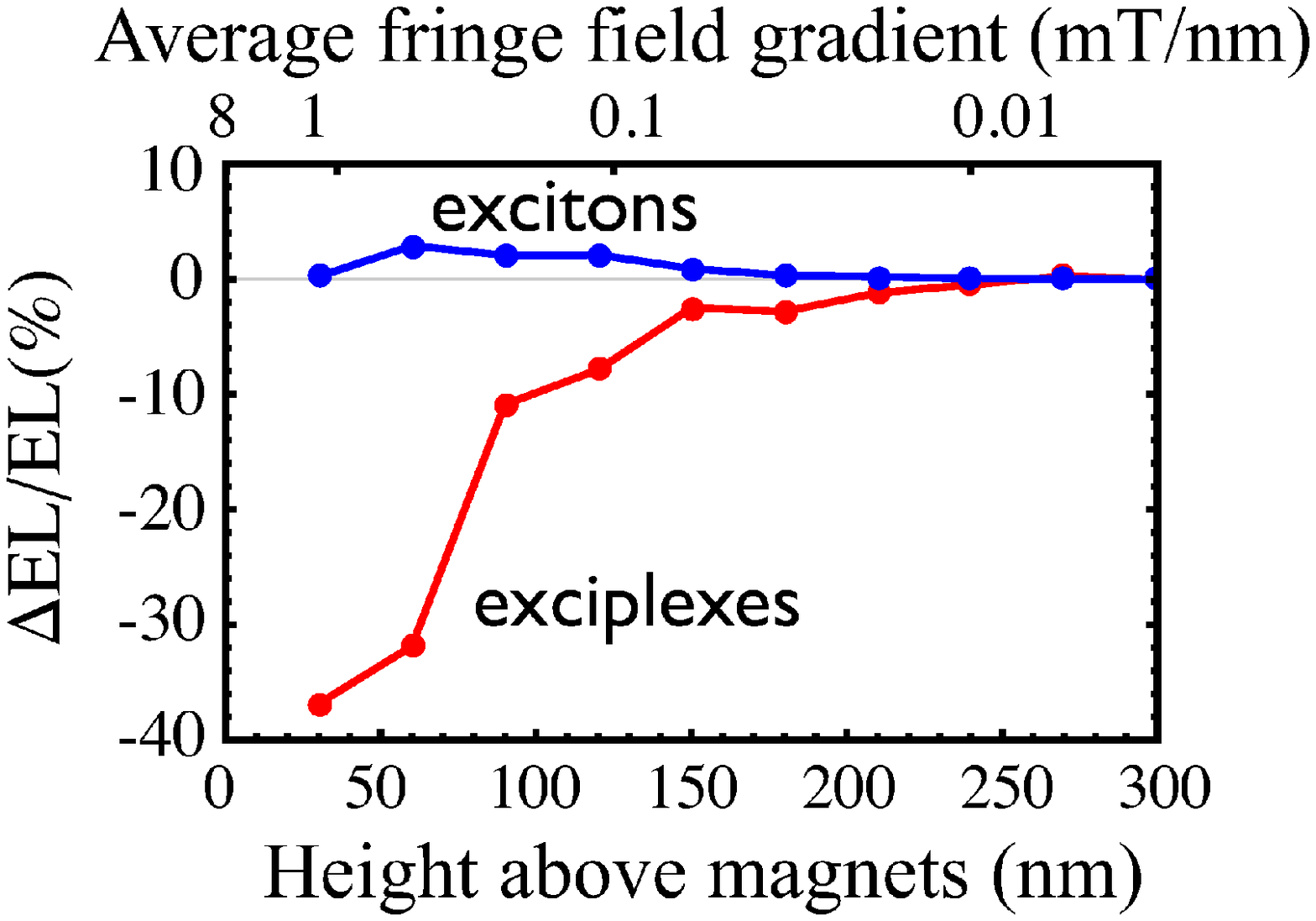}
        \caption[]
{
Exciplexes (red): $k_S = 3 \times 10^{-3}$ ns$^{-1}$, $k_T = 0$, $k_D = 10^{-6}$ ns$^{-1}$.
Excitons (blue): $k_S = 6 \times 10^{-1}$ ns$^{-1}$, $k_T = 12 \times 10^{-1}$ ns$^{-1}$, $k_D = 0$.\cite{Liu2014}
The thickness of the organic layer is taken to be 10 nm.
Shorter heights are less valid since the condition $t << Z_{min}$ is not met for Eq. \ref{eq:approx}.
}\label{fig:FF+HF}
        \end{centering}
\end{figure}

Figure \ref{fig:FF+HF} shows the figure of merit when HF is included as well. 
Given our definition of the figure of merit, at small fringe field (high above the magnets), it is zero because the ON and OFF do not differ enough.
The fringe field acts much like an external field for the usual (HF-based) magnetoelectroluminescence for spin pairs which is indicated by the sign of the figure of merits which is opposite those in Fig. \ref{fig:MEL}.\cite{Kersten2011a, Wang2016} Even though HF mixing is dominating the response (i.e. not seeing the 100\% or 300\%), we see that the fringe fields still act as an ON/OFF switch.

In this Letter we have determined the influence of patterned magnetic fringe fields on light emission in two types of OLEDs. The disparate responses between excitons and exciplexes give a simple means of discerning between the two exciplex models that have been advanced. The form of the fringe field is similar to that of the so-called isotropic $\delta g$ mechanism but since the fringe field gradient is not constrained to be aligned with the fringe field, this mechanism is more efficient in converting triplet to singlets thereby increasing radiative recombination (to 300\% instead of 100\%).
Another mechanism that we expect to yield magnetic-field effects up to 300 \% is anisotropic $g$-factor spin mixing. Up to this point, the $g$ tensor has been assumed to be isotropic which leads to only $S \leftrightarrow T_0$ spin mixing. If the two sites involved in recombination are not aligned and their $g$ tensors are anisotropic, then mixing of all four states may occur when a field is applied.

In our theory we have assumed that the organic layer is of the same lateral dimensions as the magnetic film and only examined the average responses. A future route for this research is to consider laterally small organic layers with the aim to optically detect individual domain orientations which would yield a conversion of magnetic bit information to an optical output.

This work was supported by the U.S. Department of Energy, Office of Science, Office of Basic Energy Sciences, under Award No. DE-SC0014336.

%



\end{document}